\documentclass[aps,prd,onecolumn,showpacs,reprint,nofootinbib,amsmath,amssymb,floatfix,superscriptaddress,showkeys]{revtex4}
\usepackage{graphicx}
\usepackage{dcolumn}
\usepackage{bm}
\usepackage{captcont}
\usepackage{xcolor}
\usepackage{supertabular}

\usepackage{epstopdf}
\usepackage{mathtools}
\usepackage{natbib}
\usepackage{ulem}

\begin{document}

\title{Total gravitational mass of the Galactic Double Neutron Star systems: evidence for a bimodal distribution}
\affiliation{Key Laboratory of Dark Matter and Space Astronomy, Purple Mountain Observatory, Chinese Academy of Sciences, Nanjing 210008, China}
\affiliation{School of Astronomy and Space Science, University of Science and Technology of China, Hefei, Anhui 230026, China.}
\author{Yong-Jia Huang}
\affiliation{Key Laboratory of Dark Matter and Space Astronomy, Purple Mountain Observatory, Chinese Academy of Sciences, Nanjing 210008, China}
\affiliation{School of Astronomy and Space Science, University of Science and Technology of China, Hefei, Anhui 230026, China.}
\author{Jing-Liang Jiang}
\affiliation{Key Laboratory of Dark Matter and Space Astronomy, Purple Mountain Observatory, Chinese Academy of Sciences, Nanjing 210008, China}
\affiliation{School of Astronomy and Space Science, University of Science and Technology of China, Hefei, Anhui 230026, China.}
\author{Xiang Li}
\affiliation{Key Laboratory of Dark Matter and Space Astronomy, Purple Mountain Observatory, Chinese Academy of Sciences, Nanjing 210008, China}
\author{Zhi-Ping Jin}
\affiliation{Key Laboratory of Dark Matter and Space Astronomy, Purple Mountain Observatory, Chinese Academy of Sciences, Nanjing 210008, China}
\affiliation{School of Astronomy and Space Science, University of Science and Technology of China, Hefei, Anhui 230026, China.}
\author{Yi-Zhong Fan$^\ast$}
\affiliation{Key Laboratory of Dark Matter and Space Astronomy, Purple Mountain Observatory, Chinese Academy of Sciences, Nanjing 210008, China}
\affiliation{School of Astronomy and Space Science, University of Science and Technology of China, Hefei, Anhui 230026, China.}
\author{Da-Ming Wei}
\affiliation{Key Laboratory of Dark Matter and Space Astronomy, Purple Mountain Observatory, Chinese Academy of Sciences, Nanjing 210008, China}
\affiliation{School of Astronomy and Space Science, University of Science and Technology of China, Hefei, Anhui 230026, China.}

\date{\today}

\begin{abstract}
So far, in total 15 double neutron star systems (DNSs) with a reliable measurement of the total gravitational mass ($M_{\rm T}$) have been detected in the Galaxy. In this work
we study the distribution of $M_{\rm T}$. The data prefer the double Gaussian distribution over a single Gaussian distribution and  the low and high mass populations center at
$M_{\rm T}\sim 2.58M_\odot$ and $\sim 2.72M_\odot$, respectively. The progenitor stars
of GW170817 have a $M_{\rm T}=2.74^{+0.04}_{-0.01}M_\odot$, falling into the high mass population. With a local neutron star merger rate of $\sim 10^{3}~{\rm Gpc^{-3}~yr^{-1}}$, supposing the $M_{\rm T}$ of those merging neutron stars also follow the double Gaussian distribution, the upcoming runs of the advanced LIGO/Virgo will soon detect some events with a $M_{\rm T}\lesssim 2.6M_\odot$ that can effectively probe the equation of state of the neutron stars and the distribution function is expected to be reliably reconstructed in the next decade.
\end{abstract}


\maketitle

\section{Introduction} \label{sec:intro}
In the Galaxy, so far there are sixteen double neutron star (DNS) systems have been detected \citep{Lattimer2012,Swiggum2015,Martinez2017}.
Among them, two binary systems (PSR B1913+16 and PSR J0737-3039) have provided strong indirect evidence
that gravitational radiation exists and is indeed correctly described by general relativity since the decay of these two orbits are at exactly the rates predicted by Einstein's
general theory of relativity \citep{Taylor1989,Burgay2003}. Due to the non-ignorable gravitational wave radiation, some binary neutron stars will finally merger with each other.
For instance the double pulsar system PSR J0737-3039 has a merger timescale of $\tau_{\rm gw}=85$  Myr, much shorter than the Hubble time \citep{Kramer2008}. The
coalescence of DNSs inevitably produces an energetic burst of gravitational radiation, which is one of the most promising targets for
current and the proposed future gravitational wave detectors \citep{Clark1977}. The successful detection of GW170817, a gravitational event powered by the merger of
binary NSs at a redshift of $0.0097$, by advanced LIGO/Virgo directly confirms such a long-standing speculation \citep{Abbott2017PRL-ns}. The gravitational wave data yield a
total gravitational mass of $M_{\rm T}=2.74^{+0.04}_{-0.01}M_\odot$ and favor the equation of states that predict compact neutron stars \citep{Abbott2017PRL-ns}. The
electromagnetic counterpart observation data of GW170817 favors the quick collapse of the merger remnant into a black hole. With the data of GW170817 and the mass-shedding limit
assumption, very tight constraint $M_{\rm max}<2.17M_\odot$ \citep{Margalit2017,Rezzolla2017} has been reported, where $M_{\rm max}$ is the maximum gravitational mass of the non-rotating NS.  \citet{MaPX2018} show that the constraints however can be significantly loosened if the angular momentum is much lower than the Keplerian value (i.e., $j\leq 0.8j_{\rm Kep}$). Nevertheless, some specific equation of state (EoS) models, such as MPA1 and APR3, have been ruled out because the required $j/j_{\rm Kep}$ to form a black hole is too low to be realistic. These authors also demonstrate that the DNS mergers with a $M_{\rm T}\leq  2.6M_\odot$ will shed valuable light on both the EoS model and the angular momentum of the remnants. Motivated additionally by the fact that advanced LIGO/Virgo usually give a much more accurate $M_{\rm T}$ than the masses of the individual NSs, in this work we study the distribution of $M_{\rm T}$, which is different from the previous literature \citep[e.g.][]{Schwab2010,ZhangC2011,Ozel2012,Kiziltan2013} that focus on the mass distribution of individual NSs.

\section{The sample}\label{sec:radio timing}
We refer the readers to for instance \citet{Lattimer2012} and \citet{Ozel2012} for the methods of measuring the mass of the neutron stars.
In this work, for a given DNS system, we take the neutron star masses from the latest literature. As summarized in Table 1,
so far we have in total 15 DNS systems which have a reliably measured $M_{\rm T}$.
We also list the individual mass or limit, where $M_{\rm p}$ and $M_{\rm c}$ are the gravitational mass of the pulsar and the companion, respectively.
The error bars represent the measurement uncertainty with 1$\sigma$ gaussian fluctuation.
Usually, $M_{\rm T}$ is measured much more precisely than both $M_{\rm p}$ and $M_{\rm c}$.

\begin{table}[ht]
\caption{The Galactic Double Neutron Star Systems and the masses}
\begin{ruledtabular}
\begin{tabular}{lcccl}
System & $M_{\rm T}(M_\odot)$ & $M_{\rm p}(M_\odot)$ & $M_{\rm c}(M_\odot)$ & Reference \\
\hline
J1411+2551 & 2.538(22) & $<$1.62 & $>$0.92 & \citet{Martinez2017} \\
J1757-1854 & 2.73295(9) & 1.3384(9) & 1.3946(9) & \citet{Cameron2018} \\
J0453+1559 & 2.734(3) & 1.559(5) & 1.174(4) & \citet{Martinez2015} \\
J0737-3039 & 2.58708(16) & 1.3381(7) & 1.2489(7) & \citet{Kramer2006} \\
J1518+4904 & 2.7183(7) & $0.72^{+0.51}_{-0.58}$ & $2.00^{+0.58}_{-0.51}$ & \citet{Janssen2008} \\
B1534+12 & 2.678428(18) & 1.3332(10) & 1.3452(10) &  \citet{Stairs2002} \\
J1756-2251 & 2.56999(6) & 1.341(7) & 1.230(7) & \citet{Ferdman2014} \\
J1807-2500B & 2.57190(73) & 1.3655(21) & 1.2064(20) & \citet{Lynch2012} \\
J1811-1736 & 2.57(10) & $<$1.64 & $>$0.93 & \citet{Corongiu2007} \\
J1829+2456 & 2.59(2) & $<$1.64 & $>$1.26 & \citet{Champion2005} \\
J1906+0746 & 2.6134(3) & 1.291(11) & 1.322(11) & \citet{Leeuwen2015} \\
J1913+1102 & 2.875(14) & $<$1.84 & $>$1.04 & \citet{Lazarus2016} \\
B1913+16 & 2.828378(7) & 1.4398(2) & 1.3886(2) & \citet{Weisberg2010} \\
J1930-1852 & 2.59(4) & $<$1.32 & $>$1.30 & \citet{Swiggum2015} \\
B2127+11C & 2.71279(13) & 1.358(10) & 1.354(10) & \citet{Jacoby2006} \\
\end{tabular}
\end{ruledtabular}
\label{tb1}
\end{table}

\section{The distribution of the total gravitational mass of the Galactic DNS systems}

\subsection{Modeling}
\begin{figure}[h]
\centering
\includegraphics[width=0.8\textwidth]{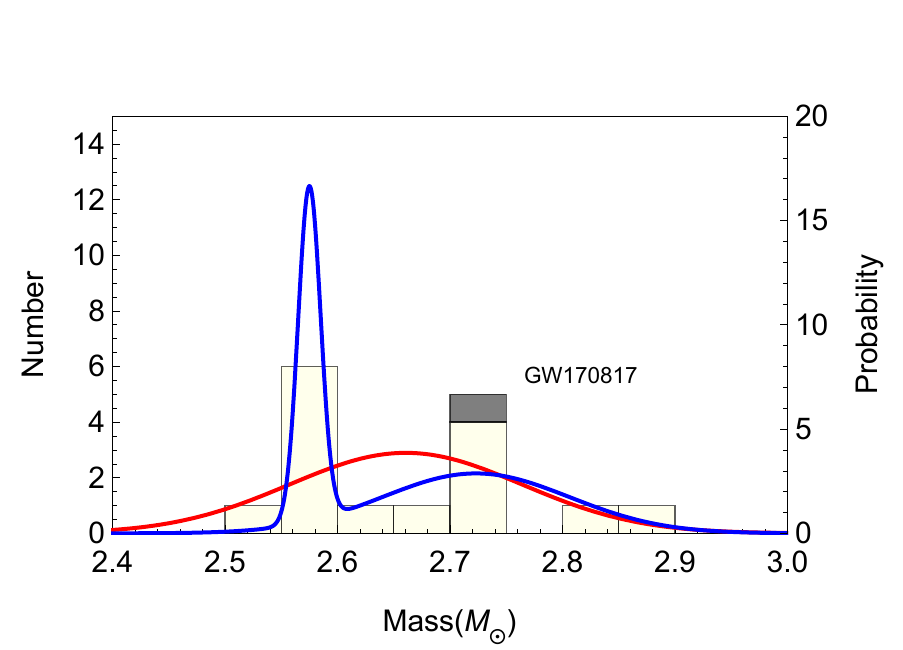}
\caption{The histogram for the total gravitational mass $(M_{\rm T})$ of 15 DNS systems and the best fit lines of the distribution probabilities. The left vertical axis represents the number of events and the right vertical axis is the probability. The red line is for the single Gaussian distribution model while the blue line represents the double Gaussian distribution model. GW170817 is among the high mass population.}
\label{fig:distribution}
\end{figure}

For the ten DNS systems summarized in \citet{Lattimer2012}, \citet{Fan2013PRD} noticed that five systems have a $M_{\rm T}\lesssim 2.6M_\odot$, and then adopted such a value to estimate the $M_{\rm max}$ under the assumption that the central engines of some short gamma-ray bursts with peculiar X-ray afterglow emission were supramassive NSs.
With the current extended sample, we have seven DNS systems with $M_{\rm T}\lesssim 2.6M_\odot$. Among them, J1411+2551 has the lowest total mass, i.e.,
$M_{\rm T}=2.538\pm0.022M_\odot$, plausibly indicating a cutoff at $\sim 2.5M_\odot$.
Fig.\ref{fig:distribution} shows the histogram of the $M_{\rm T}$ distribution, there seems one peak at $M_{\rm T}\sim 2.55-2.60M_{\odot}$, and the other
at $M_{\rm T}\sim 2.70-2.75M_{\odot}$. In the literature, the mass distribution of individual NS is usually assumed to be Gaussian \citep{Ozel2012,Kiziltan2013}. In this work we follow such assumptions. To probe possible structure in the $M_{\rm T}$ distribution, below we fit the data with a Gaussian distribution model and a double Gaussian distribution model, respectively. Note that such an approach is partially for simplicity (for example, one may also reproduce the data with a double Gaussian distribution plus an almost constant component in the mass range of $2.5-2.9M_\odot$, for which three free parameters are further introduced. With the relatively small sample, the constraints on these parameters are expected to be less tight).
The probability distribution function of Gaussian distribution model is given by
\begin{equation}
P(M_{T},M_{0},\sigma_{0}) = \frac{1}{\sqrt{2\pi}\sigma_{0}}\exp[-{\frac{(M_{T}-M_{0})^2}{2\sigma_{0}^2}}],
\end{equation}
and for the double Gaussian distribution model we have
\begin{equation}
P(M_{T},M_{1},M_{2},{\sigma_{1}},{\sigma_{2}},C) =
\frac{C}{\sqrt{2\pi}\sigma_{1}}\exp[-{\frac{(M_{T}-M_{1})^2}{2\sigma_{1}^2}}]+\frac{1-C}{\sqrt{2\pi}\sigma_{2}}\exp[-{\frac{(M_{T}-M_{2})^2}{2\sigma_{2}^2}}],
\end{equation}
where $M_{0}~(M_1,~M_2), \sigma_{0}~(\sigma_1,~\sigma_2)$ represent the mean and the variance of a Gaussian mass distribution, respectively, and $C$ is defined as a weight of the first component. In our fitting, the maximum likelihood estimate (MLE) method is adopted and the uncertainties of the measured $M_{\rm T}$ have been taken into account.
The best fit results are shown in Fig.\ref{fig:distribution} and the corresponding parameters are $(M_0,\sigma_0)=(2.67,0.10)$, $(M_1,\sigma_1)=(2.58,0.01)$, $(M_2,\sigma_2)=(2.72,0.08)$, and $C=0.40$.

To evaluate the significance of the presence of a structure in the $M_{\rm T}$ distribution (i.e., it consists of  double Gaussian components rather than a single Gaussian component) we adopt the generalized likelihood ratio test that is widely used in hypothesis testing. Our null hypothesis is the single Gaussian distribution, and the alternative hypothesis is a double Gaussian $M_{\rm T}$ distribution. We construct the likelihood ratio first, as the mass measurement of each neutron star is independent of others. The likelihood function is simply the product of probability for each independent measured total mass, i.e.,
\begin{equation}
L(M_{i}, \sigma_{i} \, | \, x, {\cal M}) = \prod\limits_{j}P(x_{j} \, | \, M_{i}, \sigma_{i}; {\cal M}),
\end{equation}
where $M_{i}, \sigma_{i}$ are the fitting parameters, ${\cal M}$ represents the model we used, and $x_{j}$ represents the data. And the likelihood ratio in our work is defined as the ratio
of the likelihood functions for two different hypotheses
\begin{equation}
\Lambda = \frac{\prod_{i = 1}^{15}P(x_{i} \, | \, M_{0}, \sigma_{0}; H_{0})}{\prod_{i = 1}^{15}P(x_{i} \, | \, M_{1}, \sigma_{1}, M_{2}, \sigma_{2}, C; H_{1})},
\end{equation}
where $x_{i}$ represents the $M_{\rm T}$ of one given DNS binary, $H_{0}$ and $H_{1}$ represent the null hypothesis and the alternative hypothesis, respectively. $M$ and $\sigma$
are determined by maximum likelihood method.
For our current data, we have
\[
\Lambda=0.011,
\]
which disfavors the null hypothesis and supports the double Gaussian distribution model. Furthermore, the test statistics $-2\ln \Lambda$ here should follow the $\chi^{2}$ distribution with three degrees of freedom \citep{Wilks1938}. We calculate the probability of the statistics, which is 97\%, corresponding to a significance level that is above $2\sigma$, suggesting that the double Gaussian distribution model is much more consistent with the data than the single Gaussian distribution model.

\section{The prospect of testing a double Gaussian distribution of $M_{\rm T}$ with advanced LIGO/Virgo}
\begin{figure}[h]
\centering
\includegraphics[width=0.8\textwidth]{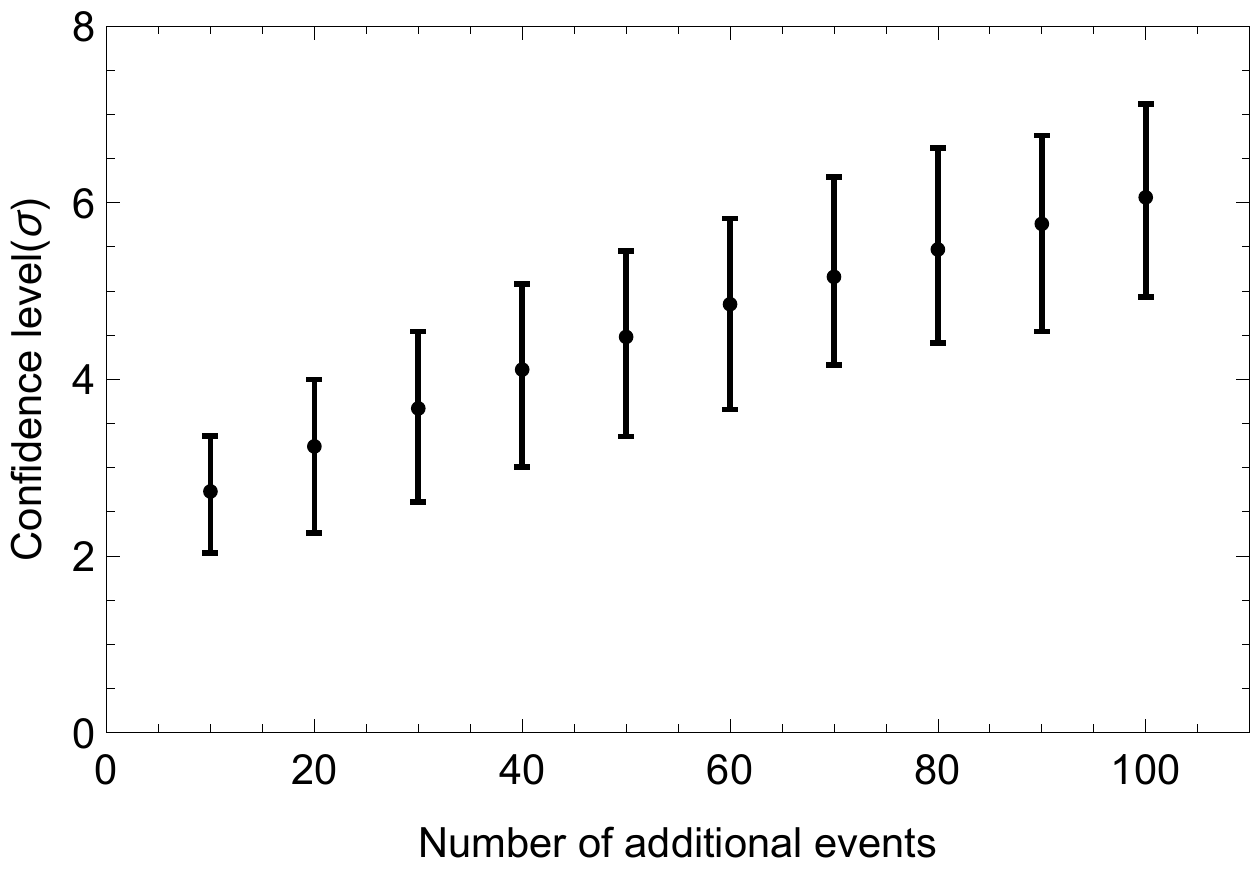}
\caption{The prospect to reliably re-construct the $M_{\rm T}$ distribution with a group of additional DNS merger events detected by advanced LIGO/Virgo, supposing these cosmological events follow the distribution shown in the blue line in Fig.\ref{fig:distribution}.}
\label{fig:projection}
\end{figure}

\begin{figure}[h]
\centering
\includegraphics[width=0.8\textwidth]{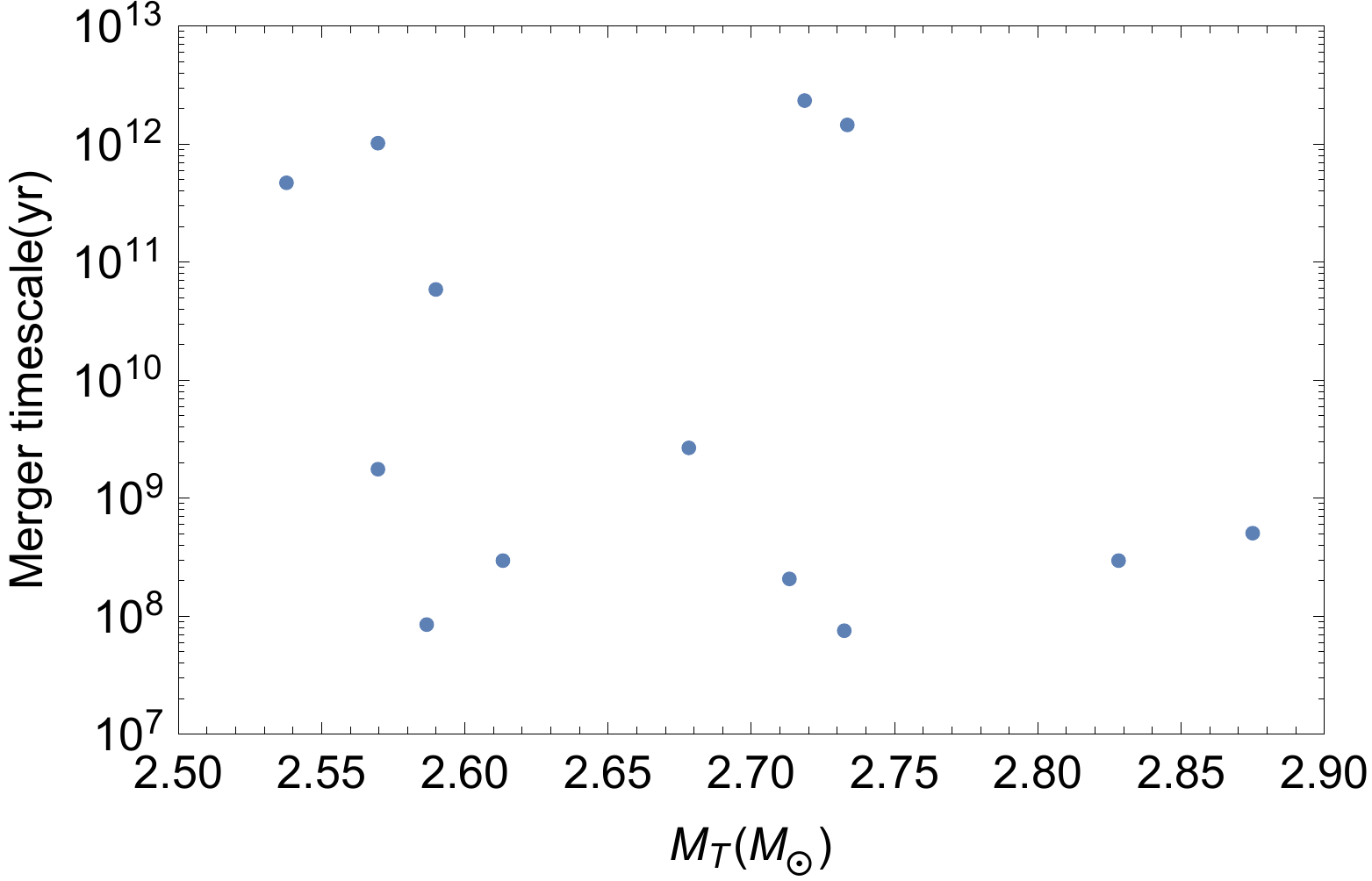}
\caption{The merger timescale versus $M_{\rm T}$ of 13 Galactic DNS systems. The merger timescales are adopted from Ref.\citep{Martinez2017,Cameron2018,Martinez2015,Kim2010,Andreoni2017,Faulkner2005,Lorimer2006,Curran1995,Nice1996,Lyne2000,Champion2004,Lazarus2016,Prince1991}.}
\label{fig:tau}
\end{figure}

The advanced LIGO/Virgo detectors can detect the gravitational wave radiation from double neutron star mergers within a typical distance $D\sim 220$ Mpc at its designed sensitivity \citep{Abadie2010}. With a DNS merger rate ${\cal R}\sim 10^{3}~{\rm Gpc^{-3}~yr^{-1}}$, as inferred from both the current gravitational wave data \citep{Abbott2017PRL-ns} and the local short GRB data \citep{Jin2018}, the detection rate of DNS mergers is thus
${R}_{\rm gw}=4\pi D^{3}{\cal R}/3\sim 45~{\rm yr^{-1}}~({\cal R}/10^{3}~{\rm Gpc^{-3}~yr^{-1}})$.
Therefore, in the next decade hundreds of DNS mergers will be detected and some of them will have accurately measured $M_{\rm T}$, with which the bimodal distribution shown in the current Galactic DNS sample can be reliably tested. As a projection, we have simulated a group of DNS merger events and examine how robust a double Gaussian distribution of $M_{\rm T}$ can be re-constructed. We combined the simulated data with 15 DNS systems in the Galaxy and the event GW170817, and the results are reported in Fig.\ref{fig:projection}. With $\sim 100$ DNS merger events detected in the future, a double Gaussian distribution of $M_{\rm T}$ can be unambiguously determined if the cosmological merging events indeed follow the total mass distribution of the Galactic DNS systems. Such an assumption may be reasonable since
the merger timescale (i.e., $\tau_{\rm gw}$) is not sensitively dependent of the masses of the neutron stars in the binary (note that the distributions of $M_{\rm p}$, $M_{\rm c}$ and $M_{\rm T}$ are actually narrow). This speculation is directly supported by the merger timescales of 13 DNS systems in the Galaxy. As shown in Fig.\ref{fig:tau},
we have $\tau_{\rm gw} \sim 10^{8}-10^{12}$ years, which spans in a huge range and the correlation coefficient is just $0.027$, which indicates no sensitive correlation between $\tau_{\rm gw}$ and $M_{\rm T}$.

\section{Discussion and Summary}
Up to now, sixteen double neutron star systems have been detected and 15 of them have a reliable measurement of the total gravitational mass $(M_{\rm T})$. These data favor
the double Gaussian distribution over a single Gaussian distribution. The best fit results are
$(M_{\rm T},\sigma)=(2.58,~0.01)M_\odot$  and $(2.72,~0.08)M_\odot$ for the low and high mass populations, respectively. The interpretation of a bimodal
distribution of $M_{\rm T}$ is somewhat challenging. One possibility is that these two populations of binary systems have suffer from different material accretion.
For instance the progenitors of the low mass population may die almost simultaneously and the material  accretion from the companion star is ignorable, while for the high mass population the death of the progenitor stars were separated and the accretion is essential. The other potential possibility is that the progenitor stars of these two populations of the DNSs have different stellar metallicities. \citet{Fryer2015} carried out the population synthesis investigation of the double neutron star binaries for two metallicities, including $Z=0.02$ (i.e., the solar composition) and $Z=0.002$ (i.e., the low metallicity model),  and found out that the low metallicity model did produce more massive neutron stars than in the solar composition scenario. Though the underlying physics/astrophysics is still to be better understood, the presence of structure in the $M_{\rm T}$ distribution can be directly tested in the next decade. Interestingly, the progenitor stars of GW170817, the first gravitational wave event powered by the merger of double neutron stars, have a $M_{\rm T}\approx 2.74~M_\odot$, which is a member of the high mass population. If the cosmological systems of the merging neutron stars also follow the $M_{\rm T}$ distribution found in this work, the upcoming O3 and the full-sensitivity runs of the advanced LIGO/Virgo will detect some events with a $M_{\rm T}\lesssim 2.6M_\odot$. As shown in
\citet{MaPX2018} the statistical study of these events would shed valuable light on possible significant rotational kinetic energy loss via high frequency gravitational waves and thermal neutrinos if supramassive neutron stars are absent. If instead some of the remnants are found to be supramassive neutron stars or even stable neutron stars, the maximal mass of the non-rotating neutron stars can be reasonably inferred. With a high neutron star merger rate (${\cal R}\sim 10^{3}~{\rm Gpc^{-3}~yr^{-1}}$) inferred from both the gravitational wave observations and the local short GRB data, the advanced LIGO/Virgo will also unambiguously test whether a double Gaussian distribution model of $M_{\rm T}$ is ``universal"  or not (or in a more general way, the $M_{\rm T}$ is structured or not).

\acknowledgments  This work was supported in part by 973 Programme of China (No. 2014CB845800), by NSFC under grants 11525313 (the National Natural Fund for Distinguished Young Scholars), 11273063, 11773078 and 11433009, by the Chinese Academy of Sciences via the Strategic Priority Research Program (No. XDB09000000) and the External Cooperation Program of BIC (No. 114332KYSB20160007).

$^\ast$Corresponding author (yzfan@pmo.ac.cn).

\end{document}